\documentclass [11pt]{article}
\usepackage{amsthm}
\usepackage{amssymb}
\usepackage{epsfig}

\newcounter{contador}
\newtheorem{propo}[contador]{Proposition}

\newcommand{\rec}{\noindent}    
\newcommand{\dem}{\rec {\it Proof. }}  

\renewcommand{\qed}{\ \hfill\rule[-1mm]{2mm}{3.2mm}}
\newcommand{\dps}{\displaystyle} 

\newcommand{\enya}{${\rm \tilde{n}}$}
\newcommand{\bx}{{\mathbf x}}

\newcommand{\K}{{\mathbb K}}
\newcommand{\R}{{\mathbb R}}
\newcommand{\C}{{\mathbb C}}
\newcommand{\N}{{\mathbb N}}

\newcommand{\U}{{\cal{U}}}

\newcommand{\RR}{{\cal{R}}}

\title{Complete integrable systems with unconfined
singularities}

\author{V\'{\i}ctor Ma\~{n}osa\footnote{Corresponding Author:
Phone: 00-34-93-727-8254; Fax:  00-34-93-739-8225.}
\\*[-.25truecm] {\small \textsl{Departament de Matem\`{a}tica Aplicada III,}}
\\*[-.25truecm] {\small \textsl{Control, Dynamics and Applications Group (CoDALab)}}
\\*[-.25truecm] {\small \textsl{Universitat Polit\`{e}cnica de Catalunya}}
\\*[-.25truecm] {\small \textsl{Colom 1, 08222 Terrassa, Spain}}
\\*[-.25truecm] {\small \textsl{victor.manosa@upc.edu}}}

\date{April 10th, 2007}
\begin{document}

\maketitle
\begin{abstract} We prove that any globally periodic rational
discrete system in $\K^k$ (where $\K$ denotes either $\R$ or
$\C$,) has unconfined singularities, zero algebraic entropy and it
is complete integrable (that is, it has as many functionally
independent first integrals as the dimension of the phase space).
In fact, for some of these systems the unconfined singularities
are the key to obtain first integrals using the Darboux-type
method of integrability.
\end{abstract}

\rec {\sl PACS numbers} 02.30.Ik, 02.30.Ks, 05.45.-a, 02.90.+p,
45.05.+x.

 \rec {\sl Keywords:} Singularity
confinement, first integrals, globally periodic discrete systems,
complete integrable discrete systems,  discrete Darboux--type
integrability method.\newline

Singularity confinement property in integrable  discrete systems
was first observed by Grammaticos, Ramani and Papageorgiou in
\cite{GRP}, when studying the propagation of singularities in the
lattice KdV equation
$x_j^{i+1}=x_{j+1}^{i-1}+{{1}/{x_j^i}}-{{1}/{x_{j+1}^i}}$, an
soon it was adopted  as a detector of integrability, and a
discrete analogous to the Painlev\'e property (see \cite{GR1,GR2}
and references therein). It is well known that some celebrated
discrete dynamical systems (DDS from now on) like the McMillan
mapping and all the discrete Painlev\' e equations satisfy the
singularity confinement property \cite{GRP,RGH}.  In \cite[p.
152]{OTGR} the authors write: ``Thus singularity confinement
appeared as a necessary condition for discrete integrability.
However the sufficiency of the criterion was not unambiguously
established''. Indeed, numerical chaos has been detected in maps
satisfying the singularity confinement property \cite{HV}. So it
is common knowledge that singularity confinement is not a
sufficient condition for integrability, and some complementary
conditions, like the algebraic entropy criterion have been
proposed to ensure sufficiency \cite{BV,LRGOT}.

On the other hand a DDS can have a first integral and do not
satisfy the singularity confinement property, as shown in the
following example given in \cite{LG}: Indeed,  consider the DDS
generated by the map $F(x,y)=(y,y^2/x)$  which has the
\textsl{first integral} given by $I(x,y)=y/x$. Recall that  a
first integral for a map $F:{\rm dom}(F)\in\K^k\rightarrow \K^k$,
is a $\K$--valued function $H$ defined in $\U$, an open subset of
${\rm dom}(F)\in\K^k$, satisfying $ H(F(\bx))=H(\bx)$ for all
$\bx\in\U.$

The above  example shows that singularity confinement is not a
necessary condition for integrability  if ``integrability'' means
the existence of a first integral. The first objective of this
letter is to point out that  more strong examples can be
constructed if there are considerd globally periodic analytic
maps. A map $F:\U\subseteq\K^k\rightarrow\U$ is globally
$p$--periodic if $F^p\equiv {\rm Id}$ in $\U$. Global periodicity
is a current issue of research see a large list of references in
\cite{CGM05,CGM}.

Indeed there exist globally periodic maps with unconfined
singularities, since global periodicity forces the singularity to
emerge after a complete period. However from \cite[Th.7]{CGM05}
it is know that
 an analytic and injective map
$F:{\cal{U}}\subseteq\K^k\rightarrow{\cal{U}}$ is globally
periodic if and only if it is \textsl{complete integrable}, that
is, there exist $k$ functionally independent analytic first in
${\cal U}$). Note that there is a difference between the
definition of complete integrable DDS  and the definition of
complete integrable continuous DS: For the later case the number
of functionally independent first integrals has to be just $k-1$,
which is the maximum possible number; see \cite{G}. This is
because the foliation induced by the $k-1$ functionally
independent first integrals generically have dimension $1$ (so
this fully determines the orbits of the flow). Hence, to fully
determine the orbits of a DDS, the foliation induced by the first
integrals must have dimension $0,$ {\it i.e.} it has to be
reduced to a set of points, so we need an extra first integral.

In this letter we only want to remark that there exist complete
integrable rational maps with unconfined singularities and zero
algebraic entropy (Proposition 1), and that these unconfined
singularities and its pre--images (the \textsl{forbidden set}) in
fact play a role in the construction of first integrals
(Proposition 2) for some globally periodic rational maps. Prior to
state this result we recall some definitions. In the following
$F$ will denote a rational map.

Given $F:{\cal{U}}\subseteq\K^k\rightarrow{\cal{U}}$, with
$F=(F_1,\ldots,F_k)$, a rational map, denote by
$${\cal S}(F)=\{
\bx\in\K^k\mbox{ such that } {\rm den}(F_i)=0\mbox{ for some }
i\in\{1,\ldots,k\}\},$$  the  {\sl singular set of $F$}. A
\textsl{singularity} for the discrete system $\bx_{n+1}=F(\bx_n)$
is a point $\bx_*\in {\cal S}(F)$. The set
$$
\Lambda(F)=\{\bx\in\K^k\mbox{ such that there exists }
n=n(\bx)\geq 1\mbox{ for which } F^{n}(\bx)\in{\cal S}(\bx)\},
$$
is called the \textsl{forbidden set} of $F$, and it is conformed
by the set of the preimages of the singular set. If $F$ is
globally periodic, then it is bijective on the \textsl{good set}
of $F$, that is ${\cal G}=\K^k\setminus\Lambda(F)$  (see
\cite{CGM} for instance). Moreover ${\cal G}$ is an open  full
measured set (\cite{R}).

 A singularity is said to be
\textsl{confined} if there exists $n_0=n_0(\bx_*)\in\N$ such that
$\lim\limits_{\bx\to \bx_*}F^{n_0}(\bx)$ exists and does not
belong to $\Lambda(F)$. This last conditions is sometimes skipped
in the literature, but if it is not included the ``confined''
singularity could re--emerge after some steps, thus really being
unconfined.

Rational maps on $\K^k$ extend to homogeneous polynomial maps on
$\K P^k$, acting on homogeneous coordinates. For instance, the
Lyness' Map $F(x,y)=(y,(a+y)/x)$, associated to celebrated Lyness'
difference equation $x_{n+2}=(a+x_{n+1})/x_n$, extends to $\K P^2$
by $F_p[x,y,z]=[xy,az^2+yz,xz]$. Let $d_n$ denote the degree of
the $n$--th iterate of the extended map once all common factors
have been removed. According to \cite{BV}, the algebraic entropy
of $F$ is defined by $E(F)=\lim\limits_{n\to\infty}\log{(d_n)}/n.$

The first result of the paper is

\begin{propo}\label{main}
Let $F:{\cal{G}}\subseteq\K^k\rightarrow{\cal{G}}$ be a globally
$p$--periodic periodic rational map. Then the following
statements hold.

\noindent (a) $F$ has $k$ functionally independent rational first
integrals (complete integrability).

\noindent (b) All the singularities are unconfined.

\noindent (c) The algebraic entropy of $F$ is zero.
\end{propo}

\noindent \textbf{Proof.} Statement (a) is a direct consequence
of \cite[Th.7]{CGM05}  whose proof indicates how to construct $k$
rational first integrals using symmetric polynomials as
generating functions.

(b) Let $\bx_*\in S(F)$, be a confined singularity of $F$ (that
is, there exists $n_0\in \N$ such that
$\bx_{\{n_0,*\}}:=\lim_{\bx\to\bx_*} F^{n_0}(\bx_*)$ exists and
$\bx_{\{n_0,*\}}\notin\Lambda(F)$). Consider $\epsilon\simeq {\bf
0}\in \K^k$, such that $\bx_*+\epsilon\notin \Lambda(F)$ (so that
it's periodic orbit is well defined). Set
$\bx_{\{n_0,*,\epsilon\}}:=F^{n_0}(\bx_*+\epsilon)$. The global
periodicity in $\K^k\setminus\Lambda(F)$ implies that there exists
$l\in\N$ such that
$F^{lp-n_0}(\bx_{\{n_0,*,\epsilon\}})=F^{lp}(\bx_*+\epsilon)=\bx_*+\epsilon$,
hence
$$
\lim\limits_{\epsilon\to{\bf 0}}
F^{lp-n_0}(\bx_{\{n_0,*,\epsilon\}})=\lim\limits_{\epsilon\to{\bf
0}} \bx_*+\epsilon=\bx_*.
$$
But on the other hand $$\lim\limits_{\epsilon\to{\bf 0}}
F^{lp-n_0}(\bx_{\{n_0,*,\epsilon\}})=F^{lp-n_0}(\bx_{\{n_0,*\}}).$$
Therefore $\bx_{\{n_0,*\}}\in\Lambda(F)$, which is a
contradiction.

(c) Let $\bar{F}$ denote the extension of $F$ to $\K P^k$.
$\bar{F}$ is $p$--periodic except on the set of pre--images of
$[0,\ldots,0]$ (which is not a point of $\K P^k$), hence
$d_{n+p}=d_n$ for all $n\in \N$ (where  $d_n$ denote the degree
of the $n$--th iterate once all factors have been removed).
Therefore $E(F)=\lim_{n\to\infty} \log{(d_n)}/n=0$.\qed

 As an example, consider for instance the globally $5$--periodic
 map $F(x,y)=(y,(1+y)/x)$, associated to the Lyness' difference equation
 $x_{n+2}=(1+x_{n+1})/x_n$, which is posses  the unconfined
singularity pattern $\{0,1\infty,\infty,1\}$. Indeed, consider an
initial condition $\bx_0=(\varepsilon,y)$ with $|\varepsilon|\ll
1$, and $y\neq -1$, $y\neq 0$ and $1+y+\varepsilon\neq 0$ (that
is, close enough to the singularity, but neither in the ${\cal
S}(F) $ nor in $\Lambda(F)$). Then
$\bx_1=F(\bx_0)=(y,(1+y)/\varepsilon)$,
$\bx_2=F(\bx_1)=((1+y)/\varepsilon,(1+y+\varepsilon)/(\varepsilon
y))$, $\bx_3=F(\bx_2)=((1+y+\varepsilon)/(\varepsilon
y),(1+\varepsilon)/y)$, and
$\bx_4=F(\bx_3)=((1+\varepsilon)/y,\varepsilon)$,  and finally
$\bx_5=F(\bx_4)=\bx_0$. Therefore the singularity is unconfined
since it propagates indefinitely.


 But the Lyness' equation
is complete integrable since it has the following two functionally
independent first integrals \cite{CGM05}:
$$\begin{array}{rl}
H(x,y)&=\displaystyle{\frac{x{y}^{4}+ p_3(x){y}^{3}+ p_2(x)
{y}^{2}+
p_1(x)y+p_0(x)}{x^2y^2}},\\
I(x,y)&=\displaystyle{\frac{(1+x)(1+y)(1+x+y)}{xy}}.
\end{array}$$
Where $p_0(x)={x}^{3}+2{x}^{2}+x$, $p_1(x)=
{x}^{4}+2{x}^{3}+3{x}^{2}+3x +1$, $p_2(x)={x}^{3}+ 5{x}^{2}+3x+2$,
$p_3(x)={x}^{3}+{x}^{2}+2x+1$. The extension of $F$ to $\C P^2$
is given by $\bar{F}[x,y,z]=[xy,z(y+z),xz]$, which is again
$5$--periodic, hence $d_n=d_{n+5}$ for all $n\in \N$, and the
algebraic entropy is $E(F)=\lim_{n\to\infty}\log{(d_n)}/n=0$.

More examples of systems with complete integrability, zero
algebraic entropy and unconfined singularities, together with the
complete set of first integrals can be found in \cite{CGM05}.

The second objective of this letter is to notice that the
unconfined singularities can even play an essential role in order
to construct a Darbouxian--type first integral of some DDS, since
they can help to obtain a \textsl{closed set of functions} for
their associated maps. This is the case of \textsl{some}  rational
globally periodic difference equations, for instance the ones
given by
\begin{equation}\label{totes}\begin{array}{c}
x_{n+2}=\dps{\frac{1+x_{n+1}}{x_n}},
x_{n+3}=\dps{\frac{1+x_{n+1}+x_{n+2}}{x_n}}, \mbox{ and }
x_{n+3}=\dps{\frac{-1+x_{n+1}-x_{n+2}}{x_n}},
\end{array}
\end{equation}
 To show this role we apply
the Darboux--type method of integrability for DDS (developed in
\cite{GM} and \cite[Appendix]{dyn-cgm}) to find first integrals
for maps.

Set $F:{\cal G}\subseteq\K^k\rightarrow\K^k$. Recall that a set
of functions $\RR=\{R_i\}_{i\in\{1,\dots,m\}}$ is \textsl{closed
under} $F$ if for all $i\in\{1,\dots,m\}$, there exist functions
$K_i$ and constants $\alpha_{i,j}$, such that
$$
R_i(F)=K_i \left(\prod_{j=1}^m R_j^{\alpha_{i,j}}\right),
$$
with $\prod_{j=1}^m R_j^{\alpha_{i,j}}\ne1.$  Each function $K_i$
is called the \textsl{cofactor} of $R_i$. Very briefly, the method
works as follows:  If there exist a closed set of functions for
$F$, say $\RR=\{R_i\}_{i\in\{1,\dots,m\}}$, it can be tested if
the function $H(\bx)=\prod_{i=1}^m R_i^{\beta_i}(\bx)$ gives a
first integral for some values $\beta_i$, just imposing $H(F)=H$.

In this letter, we will use the  unconfined singularities of the
maps associated to equations in (\ref{totes}) and its pre--images
to generate closed set of functions.

\begin{propo}\label{intesenzilles}
Condider the maps $F_1(x,y)=$$\left(y,{(1+y)}/{x}\right)$,
$F_2(x,y,z)=$$\left(y,z,{(1+y+z)}/{x}\right)$, and
$F_3(x,y,z)=$$\left(y,z,{(-1+y-z)}/{x}\right)$ associated to
equations in (\ref{totes}) respectively. The following statements
hold:

\noindent (i) The globally $5$--periodic map $F_1$ has the closed
set of functions ${\cal{R}}_1=\{x,y,1+y,1+x+y,1+x\}$, which
describe $\Lambda(F_1)$, and  generates the first integral
$$I_1(x,y)=\displaystyle{\frac{(1+x)(1+y)(1+x+y)}{xy}}.$$

\noindent (ii) The globally $8$--periodic map $F_2$ has the closed
set of functions ${\cal{R}}_2=\{x,y,z,1+y+z,1+x+y+z+xz,1+x+y\}$,
which describe $\Lambda(F_2)$, and  generates the first integral
$$I_2(x,y,z)=\displaystyle{\frac{(1+y+z)(1+x+y)(1+x+y+z+xz)}{xyz}}.$$

\noindent (iii) The map $F_3$ has the closed set of functions
${\cal{R}}_3=\{x,y,z,-1+y-z,1-x-y+z+xz,-1+x-z-xy-xz+y^2-yz,1-x+y+z+xz,-1+x-y\}$,
which describe $\Lambda(F_3)$, and generates the first integral

\noindent
$I_3(x,y,z)=(-1+y-z)(1-x-y+z+xz)(1-x+y+z+xz)(-1+x-z-xy-xz+y^2-yz)(x-y-1)/(x^2y^2z^2).$
\end{propo} \rec  \textbf{Proof.}  We only proof statement (ii)
since statements (i) and (iii) can be obtained in  the same way.
Indeed, observe that $\{x=0\}$ is the singular set of $F_2$. We
start the process of characterizing the pre--images of the
singular set by setting $R_1=x$ as a ``candidate'' to be a factor
of a possible first integral. $R_1(F_2)=y$, so $\{y=0\}$ is a
pre--image of the singular set $\{R_1=0\}$. Set $R_2=y$, then
$R_2(F_2)=z$ in this way we can keep track of the candidates to
be factors of $I_4$. In summary:
$$
\begin{array}{lcl}
  R_1:=x &\Rightarrow & R_1(F_2)=y, \\
  R_2:=y&\Rightarrow & R_2(F_2)=z, \\
  R_3:=z & \Rightarrow& R_3(F_2)=(1+y+z)/x=(1+y+z)/R_1, \\
  R_4:=1+y+z& \Rightarrow& R_4(F_2)=(1+x+y+z+xz)/x=(1+x+y+z+xz)/R_1, \\
  R_5:=1+x+y+z+xz  &\Rightarrow & R_5(F_2)=(1+y+z)(1+x+y)/x=R_4(1+x+y)/R_1, \\
  R_6:=1+x+y&\Rightarrow & R_6(F_2)=1+y+z=R_4.
\end{array}
$$ From this computations we can observe that ${\cal
R}_2=\{R_i\}_{i=1,\dots,6}$ is a closed set under $F_2$. Hence a
natural candidate to be a first integral is $$
  I(x,y)=  x^\alpha y^\beta z^\delta
  (1+y+z)^\gamma(1+x+y)^\sigma(1+x+y+z+xz)^\tau
$$ Imposing $I(F_2)=I$, we get that $I$ is a first integral if
$\alpha=-\tau$, $\beta=-\tau$, $\delta=-\tau$, $\gamma=\tau$, and
$\sigma=\tau$. Taking $\tau=1$, we obtain $I_2$.\qed

A complete set of first integrals for the above maps can be found
in \cite{CGM05}.

As a corollary of both the method and Proposition
\ref{intesenzilles} we re--obtain the recently discovered second
first integral of the third--order Lyness' equations (also named
Todd's equation). This ``second'' invariant was already obtained
independently in \cite{CGM05} and \cite{GKI}, with other methods.
The knowledge of this second first integral has allowed some
progress in the study of the dynamics of the third order Lyness'
equation \cite{CGM06}.

\begin{propo}\label{nouinvariant}
The set of functions ${\cal R}=\{x,y,z,1+y+z,1+x+y,a+x+y+z+xz\}$
is closed under the map
$F_a(x,y,z)=\left(y,z,{(a+y+z)}/{x}\right)$ with $a\in\R$, which
is associated to the third order Lyness' equation
$x_{n+3}=(a+x_{n+1}+x_{n+2})/x_n$. And gives the first integral
$$H_a(x,y,z)=\displaystyle{\frac{(1+y+z)(1+x+y)(a+x+y+z+xz)}{xyz}}.$$
\end{propo}

\dem  Taking into account that from Proposition 2 (ii) when $a=1$,
$I_2$ is a first integral for $F_{\{a=1\}}(x,y,z)$, it seem that a
natural candidate to be a first integrals could be
$$
H_{\alpha,\beta,\gamma}(x,y,z)=\frac{(\alpha+y+z)(\beta+x+y)(\gamma+x+y+z+xz)}{xyz}
$$
for some constants $\alpha$, $\beta$ and $\gamma$. Observe that
$$
\begin{array}{lcl}
  R_1:=x &\Rightarrow & R_1(F_a)=y, \\
  R_2:=y&\Rightarrow & R_2(F_a)=z, \\
  R_3:=z & \Rightarrow& R_3(F_a)=(a+y+z)/x=K_3/R_1,\mbox{ where } K_3=a+y+z,
\end{array}
$$
at this point we stop the pursuit of the pre--images of the
singularities because they grow indefinitely, and this way
doesn't seem to be a good way to obtain a family of functions
closed under $F_a$.  But we can keep track of the rest of factors
in $H_{\alpha,\beta,\gamma}$.
$$
\begin{array}{lcl}
  R_4:=\alpha+y+z& \Rightarrow& R_4(F_a)=(a+\alpha x+y+z+xz)/x, \\
  R_5:=\beta+x+y&\Rightarrow & R_5(F_a)=\beta+y+z,\\
  R_6:=\gamma+x+y+z+xz  &\Rightarrow & R_6(F_a)=(\gamma+y+z)x+(a+y+z)(1+y)/x. \\
\end{array}$$
Observe that if  we take $\alpha=1$, $\beta=1$, and $\gamma=a$, we
obtain  $R_4(F_a)=R_6/R_1$, $R_5(F_a)=R_4$ and
$R_6(F_a)=K_3(R_5/R_1)$. Therefore $\{R_i\}_{i=1,\dots,6}$ is
closed under $F_a$, furthermore $H_a=(R_4 R_5 R_6)/(R_1 R_2 R_3)$
is such that $H_a(F_a)=H_a$\qed

In conclusion, singularity confinement is a feature which is
present in many integrable discrete systems but the existence of
complete integrable discrete systems with unconfined singularities
evidences that is not a necessary condition for integrability (at
least  when ``integrability'' means existence of at least an
invariant of motion, a first integral). However it is true that
globally periodic systems are themselves ``singular'' in the
sense that they are sparse, typically non--generic when
significant classes of DDS (like the rational ones) are
considered.

 Thus, the  large number of integrable examples satisfying the the
singularity confinement property together with the result in
\cite[p.1207]{LG} (where an extended, an not usual, notion of the
singularity confinement property must be introduced in order to
avoid the periodic singularity propagation phenomenon reported in
this letter -see the definition of \textsl{periodic
singularities} in p. 1204-) evidences that singularity
confinement still can be considered as a good heuristic indicator
of ``integrability'' and that perhaps there exists an interesting
geometric interpretation linking both properties. However,
although some alternative directions have been started (see
\cite{AHH} for instance), still a lot of research must to be done
in order to understand the role of singularities of discrete
systems, their structure and properties in relation with the
integrability issues.

\noindent {\bf Acknowledgements.} The author is  partially
supported by CICYT through grant  DPI2005-08-668-C03-01. CoDALab
group is partially supported by the Government of Catalonia
through the SGR program. The author express, as always, his deep
gratitude to A. Cima and A. Gasull for their friendship, kind
criticism, and always good advice.

\end{document}